\documentclass[usenatbib]{basi}
\usepackage[T1]{fontenc}
\usepackage[british]{babel}
\usepackage[varg]{txfonts}
\newcommand{\ie}{$i.e.,\;$}
\newcommand{\eg}{$e.g.,\;$}

\usepackage{rotating}
\usepackage{dcolumn}
\begin{document}
\title[High-z radio galaxies using GMRT survey]{Unveiling the population of high-redshift radio galaxies using centimeter GMRT survey
}
\author[Singh et al.]%
       {Veeresh Singh$^1$\thanks{email: \texttt{veeresh.singh@ias.u-psud.fr}}, Alexandre Beelen$^{1}$, Yogesh Wadadekar$^2$, Sandeep Sirothia$^2$,\\ 
{\large \normalfont Ishwara-Chandra C.H.$^2$, Aritra Basu$^2$, Alain Omont$^3$ and Kim McAlpine$^4$}\\ 
       $^1$Institut d'Astrophysique Spatiale, B$\hat{\rm a}$t. 121, Universit{\'e} Paris-Sud, 91405 Orsay Cedex, France\\
       $^2$National Centre for Radio Astrophysics, TIFR, Post Bag 3, Ganeshkhind, Pune 411007, India \\
       $^3$UPMC Univ Paris 06 and CNRS, UMR 7095, Institut d'Astrophysique de Paris, F-75014, Paris, France \\   
       $^4$Department of Physics, University of the Western Cape, Private Bag X17, Bellville 7537, South Africa}
\pubyear{2014}
\volume{00}
\pagerange{\pageref{firstpage}--\pageref{lastpage}}

\date{Received --- ; accepted ---}

\maketitle
\label{firstpage}

\begin{abstract}
Ultra Steep Spectrum (USS) radio sources are one of the efficient tracers of High Redshift Radio Galaxies (HzRGs).
To search for HzRGs candidates, we investigate properties of a large sample of faint USS sources derived from our deep 325 MHz GMRT observations 
combined with 1.4 GHz VLA data on the two subfields ({\ie}VLA-VIMOS VLT Deep Survey (VVDS) and Subaru X-ray Deep Field (SXDF)) in the XMM-LSS field.
The available redshift estimates show that majority of our USS sample sources are at higher redshifts with 
the median redshifts $\sim$ 1.18 and $\sim$ 1.57 in the VLA-VVDS and SXDF fields.
In the VLA-VVDS field, $\sim$ 20$\%$ of USS sources lack the redshift estimates as well as the detection in 
the deep optical, IR surveys, and thus these sources may be considered as potential high-z candidates. 
The radio luminosity distributions suggest that a substantial fraction ($\sim$ 40$\%$) of our USS sample sources 
are radio-loud sources, distributed over redshifts $\sim$ 0.5 to 4. 

\end{abstract}


\section{Introduction}\label{s:intro}
High-redshift radio galaxies (HzRGs) are found to be hosted in massive intensely star forming galaxies which contain large reservoir of 
dust and gas (Seymour et al. 2007). 
Also, HzRGs are often found to be associated with over-densities {\ie}proto-clusters and clusters of galaxies at redshifts $\sim$ 2 - 5 
(Galametz et al. 2012). Therefore, identification and study of HzRGs allow us to better understand the formation and evolution of galaxies 
at higher redshifts and in dense environments. 
In the literature, the well known $z-{\alpha}$ relation {\ie}the positive correlation between cosmological redshift 
and the steepness of the radio spectrum has been exploited to search HzRGs (Ker et al. 2012). 
The low$-$frequency radio observations are found to be more advantageous in detecting faint Ultra Steep Spectrum (USS) radio sources 
(in turn HzRG candidates) as their flux density is relatively higher at low-frequency due to their steeper spectral index. 
Using our deep 325 MHz low-frequency GMRT observations (5$\sigma$ $\sim$ 800 $\mu$Jy) in combination with the deep 1.4 GHz observations 
(5$\sigma$ $\sim$ 80 $-$ 100 $\mu$Jy) over $\sim$ 1.0 deg$^{-2}$ in 
the VLA $-$ VIMOS VLT Deep Survey (VVDS) field and $\sim$ 0.8 deg$^{-2}$ in the Subaru X-ray Deep Field (SXDF) field, we derive a sample of 
USS sources and investigate their nature to find HzRGs candidates at submJy flux level. 
Furthermore, it is interesting to study faint USS sources down to submJy level, 
as the radio population at submJy level appears to be different than that at brighter end {\ie}above few mJy (Smol{\v c}i{\'c} et al. 2008).

\section{USS sample}\label{s:sample}
We cross-matched 325 MHz GMRT and 1.4 GHz VLA radio sources, and obtained a total 338 and 190 radio 
sources (with $\geq$ 5$\sigma$ detection at both frequencies) in the VLA-VVDS and SXDF fields, respectively. 
The two point spectral index ($\alpha{_{\rm 325~MHz}^{\rm 1.4~GHz}}$) distribution has median of $\sim$ -0.8 with standard deviation of $\sim$ 0.4. 
We selected USS sources using spectral index cut-off limit ${\alpha}^{\rm 1.4~GHz}_{\rm 325~MHz}$ $\leq$ -1.0 ({\ie}spectra steeper than -1.0) 
which resulted in total 111 and 39 USS sources in the VLA-VVDS and SXDF fields, respectively.

\section{Redshift distributions}\label{s:redshift}
\begin{figure}
\centering
\includegraphics[angle=0,width=6.0cm,height=5.0cm]{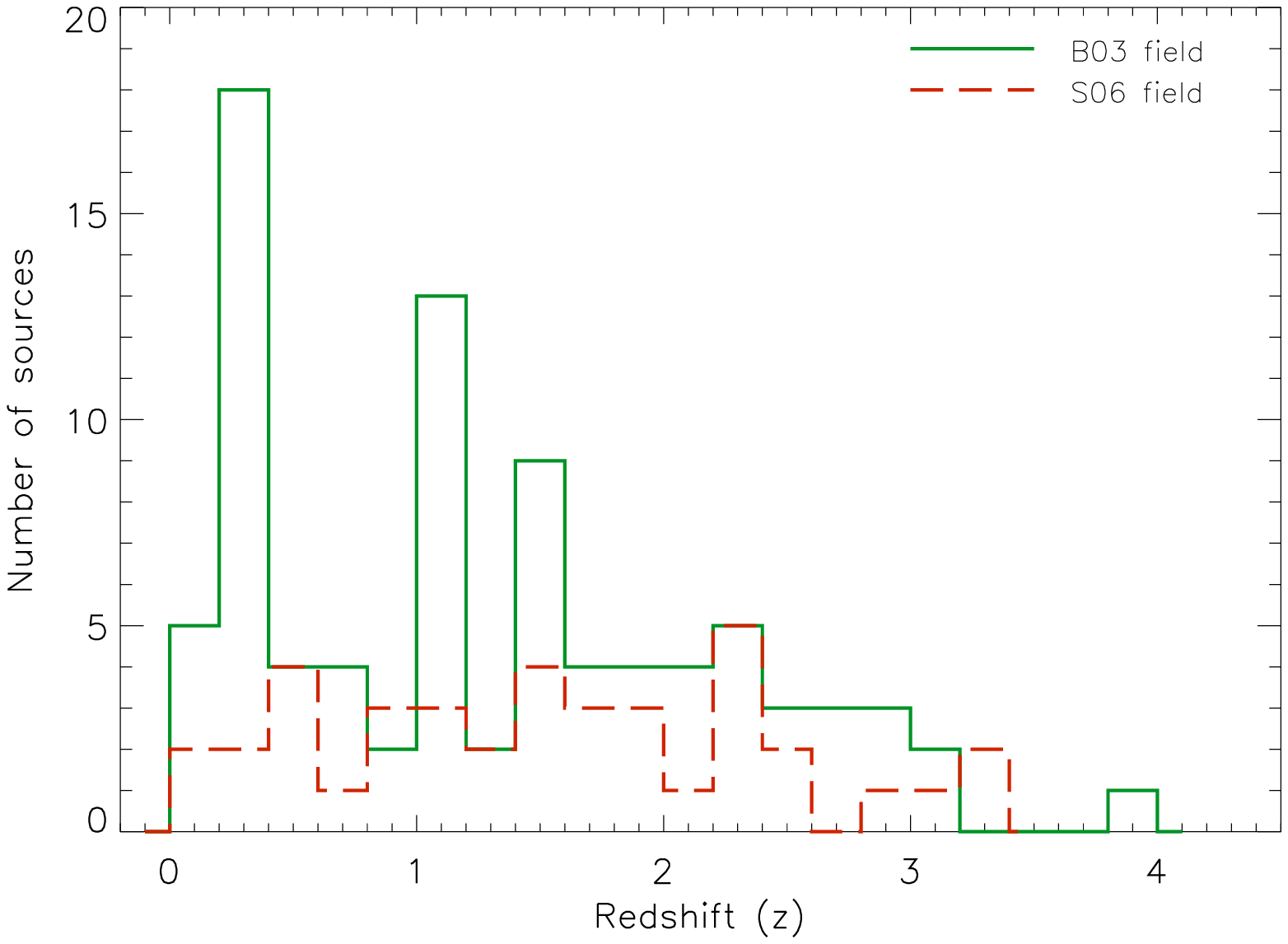}{\includegraphics[angle=0,width=6.0cm,height=5.0cm]{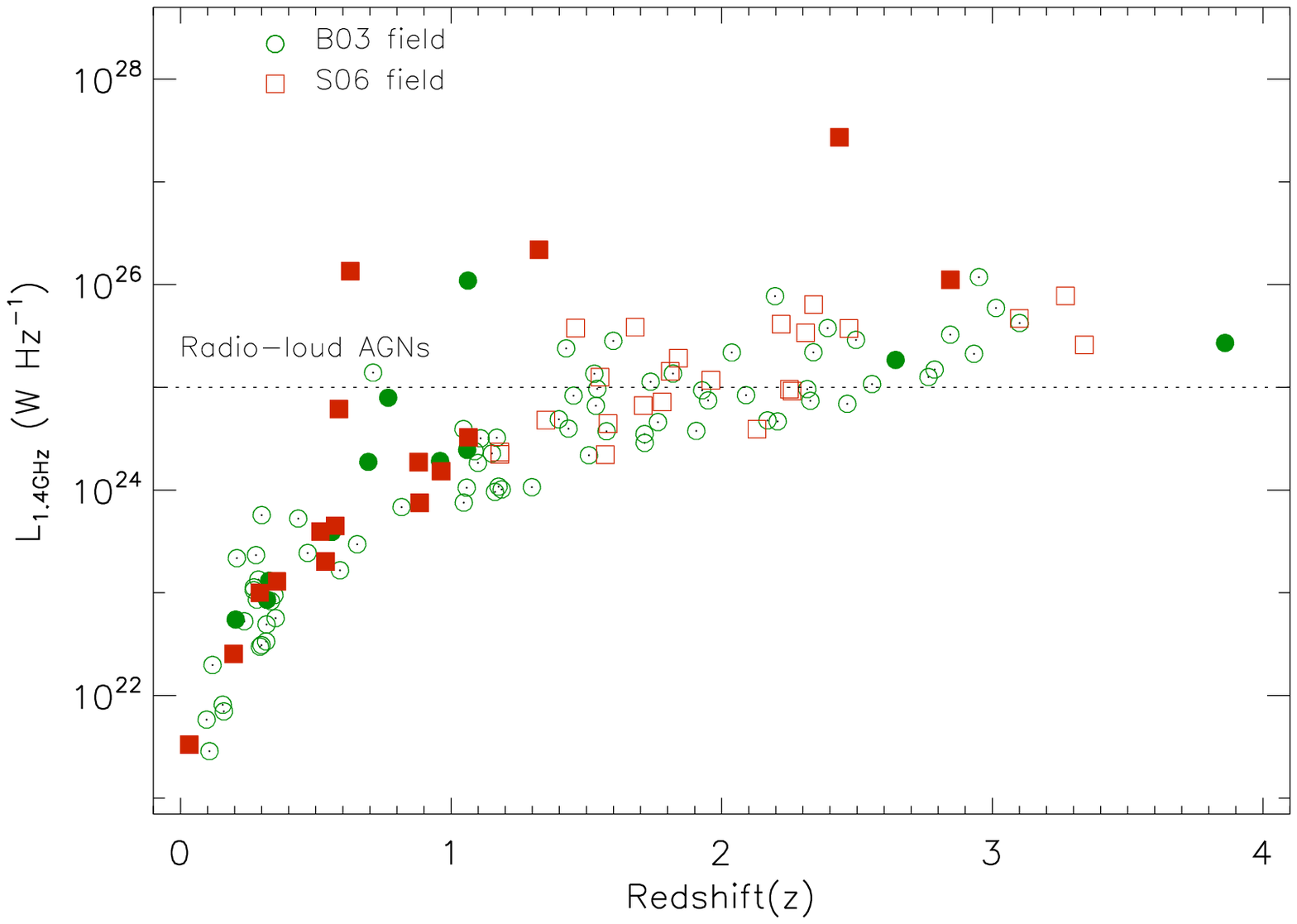}}
\caption{{\it Left}: The redshift distributions of our USS sources in the VLA-VVDS ({\ie}B03) field and in the SXDF ({\ie}S06) field. {\it Right}: 
1.4 GHz radio luminosity versus redshift plot for our USS sources in the two sub-fields. Filled and open symbols represent sources with spectroscopic and photometric redshifts, respectively. 
}
\label{fig:RedshiftHist}
\end{figure}

In the VLA-VVDS field, McAlpine et al. (2013) present photometric redshift estimates of $\sim$ 951/1054 1.4 GHz radio sources 
using 10$-$bands deep photometric data. Using photometric redshift catalog of McAlpine et al. (2013), 
we find that photometric redshift estimates are available for 86/111 USS sources 
in the VLA-VVDS field. Also, 11 of these 86 USS sources have spectroscopic redshifts measurements from VVDS spectroscopic survey. 
Rest of the 25/111 ($\sim$ 22.5$\%$) USS sources do not have redshift estimates and may be 
potential high-redshift candidates as these are faint to be detected in existing optical, IR surveys. 
In the SXDF field, Simpson et al. (2012) present spectroscopic and/or 11$-$bands photometric redshifts for 505/512 1.4 GHz radio sources. 
Using redshift estimates from Simpson et al. (2012), we find that all our 39 USS source have redshift estimates {\ie}16/39 USS sources have 
spectroscopic redshifts, while rest of the 23/39 USS sources only have photometric estimates.   
Left panel of Figure~\ref{fig:RedshiftHist} shows the redshift distributions of our USS sources in the two subfields. 
We note that the redshift distribution of USS sources in the VLA-VVDS field spans over 0.096 to 3.86 with median ($z_{\rm median}$) $\sim$ 1.18.
It is evident that a substantially large fraction (53/86 $\sim$ 61.5$\%$) of USS sources in the VLA-VVDS field are at $z \geq 1$. 
The redshift distribution of USS sources in the SXDF field is flatter and spans over 0.033 to 3.34 with $z_{\rm median}$ $\sim$ 1.57. 
We note that 27/39 $\simeq$ 69$\%$ of USS sources in the SXDF field are at redshifts ($z$) $\geq$ 1.
The lower median redshift in the VLA-VVDS field is possibly due to the fact that there are no redshift estimates 
for 25/111 ($\sim$ 22.5$\%$) USS sources in this field. 

\section{Radio luminosities of USS sources}
We study radio luminosity distributions of our USS sample sources using the K-corrected rest-frame luminosities.
Right panel of Figure~\ref{fig:RedshiftHist} shows the 1.4 GHz radio luminosity versus redshift plot for our USS sample sources. 
Most of the low-redshift ($z < 0.5$) USS sources with L$_{\rm 1.4~GHz}$ $\sim$ 10$^{21}$ $-$ 10$^{23}$ W Hz$^{-1}$ are 
likely to be radio-quiet AGNs or starforming galaxies,
However, a substantially large fraction ({\ie}55/86 $\sim$ 64$\%$ sources in the VLA-VVDS field, 
and 31/39 $\sim$ 80$\%$ sources in the SXDF field) of our USS sources do have 1.4 GHz radio luminosity higher than 10$^{24}$ W Hz$^{-1}$.
Furthermore, 22/86 $\sim$ 26.6$\%$ sources in the VLA-VVDS field, 
and 17/39 $\sim$ 44$\%$ sources in the SXDF field, do have L$_{\rm 1.4~GHz}$ $\geq$ 10$^{25}$ W Hz$^{-1}$, 
and these can be considered as secure candidates of radio loud AGNs (see Sajina et al. 2008). 
We note that some of our USS sources do clearly show double-lobe radio morphology and can be classified as FRI/FRII radio galaxies.
Majority of USS sources with relatively modest radio luminosity of L$_{\rm 1.4~GHz}$ $\sim$ 10$^{24}$ - 10$^{26}$ W Hz$^{-1}$ display 
unresolved radio morphologies with 6${^{\prime}}{^{\prime}}$.0 resolution beamsize that corresponds to 36.5 Kpc and 50.8 Kpc at $z$ = 0.5 and 2, 
respectively. The compact radio sizes with steep radio spectrum are indicative of these sources being Compact Steep Spectrum (CSS) sources. 
Some of the USS may also be Gigahertz Peaked Spectrum (GPS) sources at higher redshifts. 
Both CSS and GPS are widely thought to represent young radio sources {\ie}the start of the evolutionary path to large-scale radio galaxies 
({\eg}Fanti et al. 2009).  
In our companion paper (Singh et al. (2014) to appear in A\&A) we have shown that flux ratio of 1.4 GHz to 3.6 $\mu$m 
(S$_{\rm 1.4~GHz}$/S$_{\rm 3.6~{\mu}m}$) of majority of our USS sources is similar to the ones observed for Ultra Luminous IR Galaxies (ULIRGs) and 
Sub-Millimeter Galaxies (SMGs). Therefore, a large fraction of our USS sources are likely to be CSS, GPS like radio sources hosted in obscured environments 
of ULIRGs/SMGs. A significant fraction ($\sim$ 15$\%$) of our USS sample sources without redshift estimates also lack 3.6 $\mu$m detection and 
exhibit high ratio of 1.4 GHz to 3.6 $\mu$m (S$_{\rm 1.4~GHz}$/S$_{\rm 3.6~{\mu}m}$ $>$ 50) suggesting these sources to be HzRG candidates.       
%
\section{Conclusions}
With a large sample of faint USS sources (${\alpha}_{\rm 325~MHz}^{\rm 1.4~GHz}$ $\leq$ -1) 
derived from our deep 325 MHz GMRT observations and 1.4 GHz VLA data, we have shown that 
the criterion of using USS source remains an efficient method to select high redshift sources, even at submJy flux densities. 
The available redshift estimates of our USS sources have medians of $\sim$1.18 and $\sim$ 1.57, in the VLA-VVDS and SXDF fields, respectively.
A fraction ($\sim$ 15$\%$ $-$ 20$\%$) of USS sources without redshift estimates also lack optical, IR detections, and 
can be potential high-redshift candidates.
Our study shows that in addition to powerful HzRG candidates, faint USS population also constitutes relatively less radio 
powerful AGNs possibly hosted in obscured environments.   
%
\section*{Acknowledgements}
We gratefully acknowledge support from the Indo-French Center for the Promotion of Advanced Research (Centre Franco-Indien pour la 
Promotion de la Recherche Avance) under program no. 4404-3.
We thank the staff of GMRT who have made these observations possible. 
GMRT is run by the National Centre for Radio Astrophysics of the Tata Institute of Fundamental Research. 
%

\label{lastpage}

\begin{thebibliography}{}

\bibitem[Fanti(2009)Fanti]{2009, Astronomische Nachrichten, 330, 120}
Fanti, C. 2009, Astronomische Nachrichten, 330, 120

\bibitem[Galametz et al.(2012)Galametz et al.]{2012, ApJ, 749, 169}
Galametz, A., et al. 2012, ApJ, 749, 169

\bibitem[Ker et al. (2012)Ker et al.]{2012, MNRAS, 420, 2644}
Ker, L.M., et al., 2012, MNRAS, 420, 2644

\bibitem[McAlpine et al. (2013)McAlpine, Jarvis \& Bonfield]{2013, MNRAS, 436, 1084}
McAlpine, K., Jarvis, M. J., \& Bonfield, D. G. 2013, MNRAS, 436, 1084

\bibitem[Sajina et al. (2008)Sajina et al.]{2008, ApJ, 683, 659}
Sajina, A., et al. 2008, ApJ, 683, 659

\bibitem[Seymour et al. (2007)Seymour et al.]{2007, ApJS, 171, 353}
Seymour, N., et al. 2007, ApJS, 171, 353

\bibitem[Simpson et al. (2012)Simpson et al.]{2012, MNRAS, 421, 3060}
Simpson, C., et al. 2012, MNRAS, 421, 3060

\bibitem[Smol{\v c}i{\'c} et al. (2008)Smol{\v c}i{\'c} et al.]{2008, ApJS, 177, 14}
Smol{\v c}i{\'c}, V., et al. 2008, ApJS, 177, 14

\end{thebibliography}
\end{document}